\documentclass[twoside]{article}
\usepackage[latin1]{inputenc}
\usepackage{t1enc}
\usepackage{a4}
\usepackage{tabularx}
\usepackage{epsf}
\usepackage{graphicx}

\include{epsfig.sty}
\def\bref{\vspace{4pt}\noindent\hangindent=10mm}

\begin{document}

\setcounter{figure}{0}
\setcounter{section}{0}
\setcounter{equation}{0}

\begin{center}
{\Large\bf
Galaxy Collisions, Gas Stripping and Star\\[0.2cm]
Formation in the Evolution of Galaxies}\\[0.7cm]

Jan Palou\v s\\[0.17cm]
Astronomical Institute, Academy of Sciences of the Czech Republic\\
Bo\v cn\' \i \ II 1401, 141 31 Praha 4, Czech Republic\\
palous@ig.cas.cz
\end{center}

\vspace{0.5cm}

\begin{abstract}
\noindent{\it
A review of gravitational and hydrodynamical processes
during formation of clusters and evolution of galaxies is
given. Early, at the  advent of
N-body computer simulations, the importance of tidal fields in galaxy
encounters has been recognized.  
Orbits are crowded due to tides along spiral arms, where the star
formation is enhanced.
Low relative velocity encounters lead to
galaxy mergers. The central dominating galaxies in future clusters form 
before the clusters in a merging process in galaxy
groups. Galaxy clusters are composed 
in a hierarchical scenario due to relaxation processes between
galaxies and galaxy
groups. As soon as the overall cluster gravitational potential
is built, high speed galaxy versus galaxy encounters start to play a
role. These harassment events gradually thicken and
shorten spiral galaxy disks leading to the formation of S0 galaxies and
ellipticals. Another aspect of the high speed motion in the hot and
diluted Intracluster Medium (ICM) is the ram pressure exerted on the 
Interstellar Matter (ISM) leading to stripping of the ISM from
parent spirals. The combinations of tides and ram
pressure stripping efficiently removes the gas from
spirals, quenching the star formation in galactic disks, while
triggering it in the tidal arms and
at the leading edge of gaseous disk. Gas stripping  from disks
transports the metals to the ICM. In some cases, the gas extracted from 
the galactic disks becomes self-gravitating forming tidal dwarf galaxies.

Star formation (SF) is another important ingredient in the evolution 
of galaxies. Young
stars provide the energy, mass feedback and the metals to the ISM.
SF also drives the supersonic turbulence and triggers SF at other places in
the galaxy. Examples of supershells in the ISM, resultant from the evolution 
of stellar cluster, are given.  
Structures, supershells, filaments and sheets are also produced when
the ISM is compressed during galaxy collisions and when high velocity
clouds (HVC) fall and ram through gaseous galaxy disks.  In some cases,
the
compressed structures become self-gravitating and fragment to produce
clumps. When a low density medium is compressed, the clump masses and
sizes correspond to those of giant molecular clouds (GMC). When already
existing GMC's are compressed they are driven  to collapse forming massive
super-star clusters. The fragmentation process in the dense environment of
GMC's produces fragments of stellar
masses with the mass function having a slope similar to the stellar
Initial Mass Function.
}
\end{abstract}

\section{Introduction}

The internal evolution of isolated galaxies is the subject of many studies.
The evolution of stellar disks depends on the ratio between the radial 
component of the velocity dispersion $\sigma _R$ multiplied with the local
epicyclic frequency $\kappa $, which is an analog of pressure in the 
stellar disk, and the disk surface density $\Sigma $ multiplied with the 
constant of gravity $G$, which gives the local gravitational force in the 
flat disk. The above quantities are combined in the Toomre (1964, 1977) $Q$ 
parameter    
\begin{equation}
Q = {\sigma _R \kappa \over 3.36 G \Sigma }
\end{equation}
to evaluate the local stability of a rotating, self-gravitating disks:
for $Q < 1$ the disk is locally unstable and forms large scale deviations from
the axial symmetry - spiral arms. A purely stellar disk heats up, when 
individual stars scatter on spiral arms, and self-stabilizes. 

If the fraction of the disk kinetic energy in random motions is small enough,
the bar forms in its central part (Ostriker \& Peebles, 1973). The bar 
exchanges the angular momentum with stars, which results in radial 
redistribution of mass leading to the formation of a central mass 
concentration.
The growing central mass concentration partially dissolves the bar itself 
(Shen \& Sellwood, 2004).

A dissipative gaseous interstellar medium (ISM) is an additional 
component of the disk moving in a common gravitational field with stars. 
Energy dissipation, supersonic shocks and collisions of interstellar clouds 
reduce random motions of the ISM agitating spiral-like instabilities, which 
gradually heat the stellar disk. In an isolated spiral galaxy,
the heating by a bar and spiral arms increases the stellar velocity dispersion 
by less than a factor of 2 (Bournaud et al., 2004).   
 
The ISM complements the influence 
of the central mass concentration with viscous forces that 
shift systematically the gas flow-lines relative to the stellar orbits.
The evolution of a bar and spiral structure in an isolated galaxy, 
their growth and dissolution, is driven by a combination of the increase 
in stellar velocity dispersion due to scattering of individual stars 
on large scale deviations from the axial symmetry, by the growth of 
the central mass concentration and by the torque between the gaseous and 
stellar components (Bournaud \& Combes, 2004).   
 
Star formation, the mass recycling between stars and ISM, and the energy 
feedback are also recognized as fundamental processes. They are included in
closed box models the evolution of isolated galaxies 
(Jungwiert et al. 2001, 2004). 
Several issues are addressed, remaining as 
open question if, when, and how much the box has to be open to gas
infall, environmental effects, and galaxy major and minor mergers:
\begin{itemize}
\item Persistence of star formation. 
With the present rate, star formation would consume all the gas within
less than a few Gyr. Fresh gas supplies are needed to keep the 
star formation active. 
\item The metalicity distribution and chemical evolution of the disk. 
The G-dwarf problem requires gas infall.
\item  The existence of a thin gaseous disk, where recurrently 
the instabilities operate also needs accreted fresh gas.
\item Thin disks are destroyed in minor merger and harrasment events. They 
have to be recreated.
\item Renewal of bars driving the mass to the galactic center also requires
gas accretion.   
\end{itemize}

In this review, the tides, major, intermediate and minor mergers
events, harrasment events,
gas stripping, star formation and feedback are described. We discuss the 
relevance of these processes in the formation, and evolution of galaxies, and
in the evolution of galaxy groups and clusters. 

\section{Tides}

The importance of tides has been shown in early numerical N-body simulations
by Toomre \& Toomre (1972). Their restricted N-body simulations of an 
encounter between two spiral galaxies led to the basic features. As the 
galaxies pass by each other, tidal forces provide the stars and ISM of the 
disks with sufficient energy to escape the inner potential well. Two streams 
of disk material appear on both near and far side of the disks relative 
to the position of the peri-center. In the near side, stars and gas form 
a tidal bridge between the disks while the far side material forms long 
tidal tails. The length and the shape of the tails is a sensitive function 
of the relative velocity and of the geometry of the encounter: the most 
effective are the low relative velocity prograde encounters, where the 
spin and angular momentum vectors are aligned. In this case, the 
momentum and energy exchange in resonant orbits is most effective, 
leading to long tidal arms and a close interconnection between the central 
parts of the original disks, to merge in the future.          

The evolution of tidal debris is described by Mihos (2004). Most of the 
remnant material remains bound. It follows elliptical orbits and only a small 
fraction of the orbits are unbound. The fraction of gravitationally 
bound material
is proportional to the extend of the dark matter halo: large halos result in
less unbound material. The bound tidal debris return after some time to the 
galaxy disk in the form of gas infall and high velocity clouds, which 
may trigger the star formation. A fraction of the disk gas on unbound orbits
becomes part of the Intracluster  medium (ICM). The metalicity of the 
ICM, which is about 1/4 $Z_{\odot }$, gives an evidence of the connection 
between galaxy disks and ICM. The slow encounters are the most effective 
in driving the gas of the original disks inward to the center of the future 
galaxy produced by the merger, which triggers a nuclear starburst.

\section{Major mergers}

Computer simulations of 
low velocity encounters of spiral galaxies with similar masses 
- major merger events -
produce remnants that are in surprisingly good agreement with the observed 
shapes, density profiles and velocity distribution of the observed giant 
elliptical galaxies. The mass ratio of the progenitor disks determines the 
global properties of the remnant (Burkert \& Naab, 2003a, b).   
Giant elliptical galaxies can be subdivided into disky and boxy types
(Naab \& Burkert, 2003):
the less massive disky giant ellipticals show disk-to-bulge rations of S0 
galaxies and rotation in their outer parts. They can be understood 
as resulting from major mergers with mass ratios from 1/1 to 1/4.
The more massive boxy giant ellipticals form by 1 : 1 mergers with special
initial orientations only, which weakens the disk merger scenario as the 
possible formation mechanism. Other processes have to be added, such as the 
star formation and the energy dissipation in the gaseous components of 
progenitor disks. 

The present day giant elliptical galaxies may have been formed by
major mergers of gas-rich galaxies with a subsequent starburst,  or by
mergers of less gas-rich galaxies, which is more common in galaxy
clusters. The second scenario assumes that a fraction of stars 
forms before ellipticals in smaller gas-rich galaxies. In the
$\Lambda $CDM cosmological simulations the number of mergers varies
with z. The giant haloes of future elliptical exist already at $z = 6$
with the relatively small variation of mass at $z < 6$. Giant
ellipticals exist at redshifts $z = 3$ with about $50\%$ of stellar
mass, however, a typical cD galaxy has suffered significant merging events
even at redshifts $z < 1$ (Gao et al., 2004). Bimodal
distribution of metalicity observed in elliptical galaxies is a 
product of gas-rich mergers, when the globular clusters form in the
tidal arms (Li at al. 2004).
 
The Antennae, the colliding pair of galaxies NGC 4038/39, provide an nearby 
example of an early phase of a low velocity encounter between galaxies. 
The velocity field of Antennae galaxies,
showing the details of the interaction, has been measured by Amram et al. 
(1992).  Later, 
when the collision partners will not be distinguished any more, it should result
in a field elliptical galaxy.
Another example of a slightly more advanced ongoing merger event, 
is the Hickson compact group 31 (Amram et al. 2004). HCG 31 is a group in 
early phase of merger growing through slow and continuous acquisition of 
galaxies from the associated environment. 
One of the best examples of remnants after a recent merger event is
the giant elliptical galaxy  NGC 5018. Its inner part shows a uniform 
$\sim 3 $ Gyr old stellar population presumably produced in a merger
induced starburst (Buson et al., 2004).

\section{Intermediate and minor mergers}

Collisions of intermediate (1/5 - 1/10) mass ratio partners show the
intermediate  merger events, which are explored by Bournaud
et al. (2004). They result in a hybrid system with spiral-like
morphology and elliptical-like galaxy kinematics similar to some of the
objects identified by Chitre \& Jog (2002).

Minor mergers, i.e. merger of a galaxy with a satellite 1/10 or
less massive than the galaxy itself, are discussed in the context of
formation of thick disk and of disk globular clusters (Bekki \&
Chiba, 2002, Bertschik \& Burkert, 2003). The disk globular clusters
are formed in the high-pressure dense central region of the gas-rich
dwarf galaxy, which is compressed in a tidal interaction with the thin
disk of the spiral galaxy, and later they are dispersed in the disk
region, when the original dwarf merges with the galaxy. The
origin of thick galactic disks may be the result of the same minor
merger event producing the disk globular clusters.

\section{Evolution of galaxies in clusters}

An overabundance of spiral galaxies and an under abundance of S0 galaxies in
high-redshift clusters, when it is compared to low $z$ clusters, is a
consequences of galaxy merger events and of other environmental
effects. In $\Lambda $DCM cosmological models with $\Omega _0 = 0.4$
and $\Omega _\Lambda = 0.6$, the galaxy interactions and tidal effects
play a r\^ ole mainly at intermediate redshifts $0.5 < z < 5$  
(Gnedin, 2003). Galaxies
entering clusters with low relative velocities merge their halos with
the cluster and their subsequent dynamical evolution is due to perturbations along
their orbit. The infalling galaxy groups experience merger
events of their members. After virialization, when the galaxy velocity
in the cluster increases to a few $10^3$ km s$^{-1}$ the mergers are
suppressed and high-speed galaxy encounters and interactions with
the intracluster medium (ICM) start play a role.    
 
The ICM consists of hot ($\sim 10^7$K) diluted ($10^{-3} 
- 10^{-4}$ cm$^{-3}$) gas, which is detected in X-ray observations. 
The optical/infrared 
observations (Renzini, 1997) monitor the stellar component of galaxies, 
and also their chemical composition. 
The bulk of the cluster light is produced in bright giant elliptical galaxies 
and in galactic bulges by  intermediate and low mass stars.   
The content of Fe in 
clusters scales with their total light and the abundances of various elements 
does not seems to change from cluster to cluster  (Renzini, 2004). If the 
metals are mainly produced in SN Type Ia and SN Type II, the constancy of
metallic ratios implies the same universal ration of the two types of SN. 
It may be interpreted as a 
sign
that majority of cluster metals have been produced in the dominating giant 
galaxies, and that the stars formation holds a universal 
initial mass function
(IMF). It appears that the total mass of metals in the ICM 
supersedes the total mass of metals in stars (Renzini, 2004). The galaxies 
lose more metals to the ICM than they are able to retain.  Their total 
abundance does not correspond to the actual SN rates, the observations could
be reproduced if the SN Type Ia in ellipticals have 5 - 10 time larger rates
in the past compared to present and if the slope of the  
IMF is not too shallow: Salpeter (1955) IMF with the power law 
slope -2.35 matches the requirements.     

How the metals were transported to the ICM? Potential mechanisms are
the ICM ram pressure stripping of galactic ISMs, which may become more
effective in combination with gravitational tides, or the star
formation feedback connected to early stellar winds driven by
the starburst forming majority of the stellar galaxy itself. 
An evidence for early stellar winds serve the Lyman-break galaxies
(Pettini et al., 2003). Another possibility are late local winds due to
massive starbursts (Heckmann, 2003) or flows driven by a
declining rate of SN type Ia (Ciotti et al., 1991).
Both, gas extraction by stripping and gas ejection by winds will be 
described below. 

\section{Galaxy harrasment}

Moore et al. (1996) discuss the heating effects influencing sizes and 
profiles of dark matter haloes of galaxies in clusters: tidal heating of
haloes on eccentric orbits and impulsive heating from rapid encounters
between haloes. N-body simulations show that both processes
restrict the halo sizes. Frequent high speed galaxy
encounters in clusters, galaxy harassment, drive
the morphological transformation of galaxies in clusters. Moore et al.
(1995) show the dramatic evolution, which happened in clusters during
the recent $\sim $ 5 Gyr. Both, young clusters at $z \sim 0.4$ and old nearby
clusters have central dominant elliptical galaxies, which have been formed 
before clusters. The young
clusters are populated by many spiral galaxies, but the old, large virialised
systems, have inside of a deeper potential valley all other galaxy types,
like S0, dwarf Ellipticals, dwarf S0, dwarf Spheroidals and Ultra
Compact Dwarfs (Moore, 2003). Numerical simulations show that if the
cluster velocity dispersion is more than a few times the internal
velocity dispersion of galaxies, they do not merge in encounters. 
The observed morphological change is due to impulsive
interactions during high speed encounters restricting the dark matter
haloes and pumping the kinetic energy
into disks of spirals changing them to S0, dE, dS0, sSph, UCD galaxies.  
Hydrodynamical processes, like gas stripping, also matters,
and help with the change, particularly in relation to star-formation 
and ICM.

\section{Gas stripping}

Gunn \& Gott (1972) in a description of events during and after the
collapse, or formation of a galaxy cluster, predict the formation of hot
and diluted intracluster medium (ICM) produced in shock randomization 
of infalling gaseous debris. The interstellar medium (ISM) in a
galaxy moving through the ICM feels the ram pressure $P_r$, which is
proportional to 
\begin{equation}
P_r \sim \rho _{ICM} v^2,
\end{equation}
where $\rho _{ICM}$ is the density of the ICM and $v$ is the velocity
of the galaxy relative to the ICM. The ISM may be stripped away the
parent galaxy if at given distance
from the galactic center the ram pressure exceeds the gravitational
restoring force $F_r$ 
\begin{equation}
F_r = 2 \pi G \Sigma _s \Sigma _{ISM},
\end{equation}
where $\Sigma _s$ is the surface density of stars, $\Sigma _{ISM}$
is the surface density of the ISM and $G$ is the constant of gravity.  

The ram pressure gas stripping influences the presence of the ISM,
which may affect the star formation rates in galaxies. It is still an
open question, if the observation of Butcher \&
Oemler (1978, 1984), showing that distant clusters contain a far higher
fraction of blue star forming galaxies than their near-by
counter-parts, can be explained with the ICM ram pressure stripping
of the ISM from galaxies.

A mechanism that suppresses the star formation in local galaxy
clusters is apparently connected to the morphological transformation of 
galaxies: S0 galaxies are under-represented in distant galaxy
clusters, luminous spiral galaxies are in deficit near the centers of
local galaxy clusters (Dressler et al., 1997; Couch et al. 1998). 
This suggests that the cluster environment
removes the ISM from galaxies, suppressing star formation and 
transforming spirals to S0's. 

Numerical simulations of ram pressure gas stripping using
3-dimensional SPH/N-body code have been preformed by Abadi et al. (1999).
They confirm the predictions of Gunn \& Gott (1972) that the radius to
which the gas is removed from the parent galaxy depends on the
relation of the ram pressure to the restoring force. But in any case a
substantial part of the cold gas remains sufficiently bound to the
stellar disk. The star formation rate is reduced by a factor of 2 only,
which brings them to a conclusion that the simple ram pressure stripping
does not adequately explain the sharp decline of star formation seen
in Butcher-Oemler effect.

%%%%%%%%%%%%%%%% figure 1 %%%%%%%%%%%%%%%

\begin{figure}[tbh]
% \vglue-3.5cm
%\begin{center}
\vglue-0.5cm
 \includegraphics[width=13.5cm,angle=0]{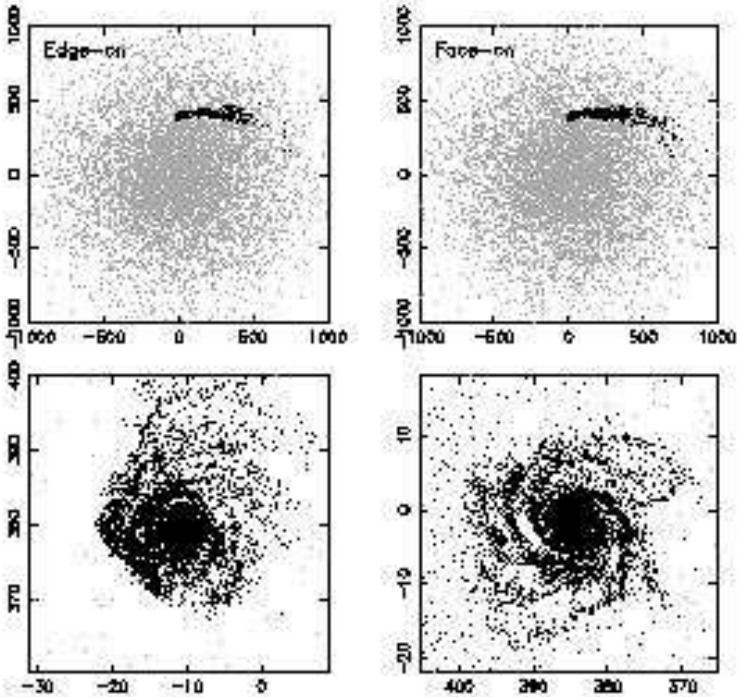}
%\label{fig1}
\caption{Simulations of ram pressure gas stripping (J\' achym,
2004). Edge-on orientation of the galaxy plane relative to the orbit
(left panel) is compared to the pole-on orientation of the galaxy
plane (right panel).
}
\label{fig11}
%\end{center}
\end{figure}

Volmer et al. (2001) show in simulations using sticky particles that
the gas stripping is rather sensitive to the galaxy orbit,
particularly to the minimum distance to the cluster center,
and also to the orientation of the galactic disk relative to
the orbit inside the cluster. In some cases the removed ISM is
re-accreted and it falls back to the galactic disk, possibly triggering
star formation in the central part of the disk within the remaining gas. 
Simulations of ram
pressure gas stripping along an orbit in a cluster by J\' achym
(2004) give similar result: the orbital parameters and the orientation
of the galaxy are important. The gas is more effectively removed when
the pole of the galactic plane is near the direction of the galaxy orbital
motion in a cluster (Fig. \ref{fig11}) compared to the situation when
the galaxy moves edge-on along its orbit.    

\section{Gas stripping and tides}

Recently reviews on the efficiency of the stripping mechanism were given
by van Gorkom (2004) and Combes (2004). The theoretical
considerations stress the effects of viscosity and
thermal conduction including Kelvin-Helmholtz instability (Nulsen,
1982). The turbulence, shells and supershells
formed by star formation increase the effect of the ram pressure
stripping due to erosion at the edges of ISM holes (Bureau \& Carignan,
2002). Ram pressure also contributes to the
overpressure inside of expanding structures increasing their 
sizes.   

A proto-type of overwhelming gas stripping in the Virgo cluster 
is the galaxy NGC 4522 with a
truncated HI disk, enhanced star formation in the central
region, extraplanar gas, while undisturbed stellar disk. The gas
distribution suggests an ICM-ISM interaction, while the undisturbed
disk rules out a gravitational interaction (Kenney \& Koopmann, 1999).  

The transformation of galaxies in dense environment results from the combined 
action of gas stripping and gravitational tides.
C 153 galaxy in the cluster Abel 2125 shows
an ongoing gas stripping: A tail of ionized gas is seen in 
[O II] emission, which extends at least 70 kpc toward the cluster
core along C 153 orbit, coincide with a soft X-ray feature seen in the Chandra
observations (Keel et al., 2004). At the same time the HST optical
picture shows clumpy morphology, including luminous star-forming
complexes and chaotic dust features. The perturbed stellar disk with
enhanced star formation activity suggests a possibility that a burst
of star formation has been initiated during the close passage of C 153
to the cluster center.  

More examples of the stripped dwarf irregular galaxies
(van Zee et al., 2004) and of galaxies with truncated star formation disks 
in Virgo cluster (Koopmann \& Kenney, 2004a, b) show both the gas stripping and
gravitational interactions including the induced star formation, however, the
dominant environmental effect on galaxies in clusters is the ram
pressure gas stripping.

\section{Tidal Dwarf Galaxies}

The long tidal tails observed in many cases of interacting galaxies show 
massive clumps of $10^9 M_{\odot }$ (e.g. IC 1182, NGC 3561,  
NGC 4676, etc.). These massive blobs has been named Tidal Dwarf Galaxies
(TDG), 
since they have galactic masses and the chemical composition corresponds to 
the recycled matter pushed out of the disks of the interacting partners.
To become a long living independent galaxy they should be gravitationally 
bound systems (Duc et al., 2000). The most prominent interacting system, 
the Antennae galaxies, has been studied by Mirabel et al. (1992), who describe a 
TDG candidate at the tip of their long tidal arm. Hibbard et al. (2001) 
also analyzed this concentration of gas and star forming regions, however,
from observations it is very difficult to assess if it is gravitationally 
bound.    

To decide if a TDG candidate will be a new galaxy formed out of an 
interaction, several questions have to be addressed (Bournaud et al. 2004):
\begin{itemize}
\item Are the blobs real concentrations in three dimensions? They may be 
just projection effects due to tidal arm geometry.
\item Are they kinematically decoupled from the tidal arms? 
Are they long living?
\item Do they contain dark matter?  
\end{itemize}
Bournaud et al. (2004) give the answer at least to the first question:
The simulations of galaxy collisions provide shapes of tidal 
arms, which may be virtually observed from all the sides. They conclude that 
some observed $10^9 M_{\odot }$ mass concentrations are real TDG candidates. 
Some of them are  self-gravitating, but to decide on their future is still 
difficult with the current resolution of simulations. The star formation, 
energy and mass feedback have to be included in the future numerical 
experiments. The third problems on the content of the dark matter also 
remains open due to the uncertainty in the internal kinematics of TDG 
candidates.

The simulations provide one more important conclusion (Duc et al. 2004): 
the existence of 
$10^9 M_{\odot }$ mass concentrations at the tips of tidal arms is rather 
sensitive to the extent and the density profile of the halo. 
It has to be extended enough ($\sim $150 kpc from the center of a 
collision partner) so that the 
collision happens within it. Then the flow lines of the perturbed gas 
from all galactocentric distances in the original disk concentrate 
at the tips of the tidal arms, giving a
kinematical origin to the TDG. When the halo is too concentrated so that the 
collision happens at the distance, where the rotation curve already 
decreases, the 
perturbed gas populates all the tidal arm and there is no place
where $10^9 M_{\odot }$ may gather.       

The kinematical gathering of stars and gas distinguishes TDG from 
super-star clusters (SSC), which are seen the interacting galaxies. 
SSC are not only less massive ($10^5 - 10^7 M_{\odot }$) but they arise from  
gravitational instabilities in the stellar or gaseous components along the 
tidal arms.

\section{SMC, LMC and the Milky Way system}
 
The Small (SMC), and Large Magellanic clouds (LMC) and the
Milky Way form the nearest interacting system of galaxies, where
the gravitational and hydrodynamical processes can be studied.
We see the result of a combination of gravitational tidal forces with
gaseous ram pressure. The star formation and the mass and energy
feedback is also involved as demonstrated with many expanding ISM shells
and supershells.

\begin{figure}
\includegraphics*[bb=80 170 545 800, width=11cm]{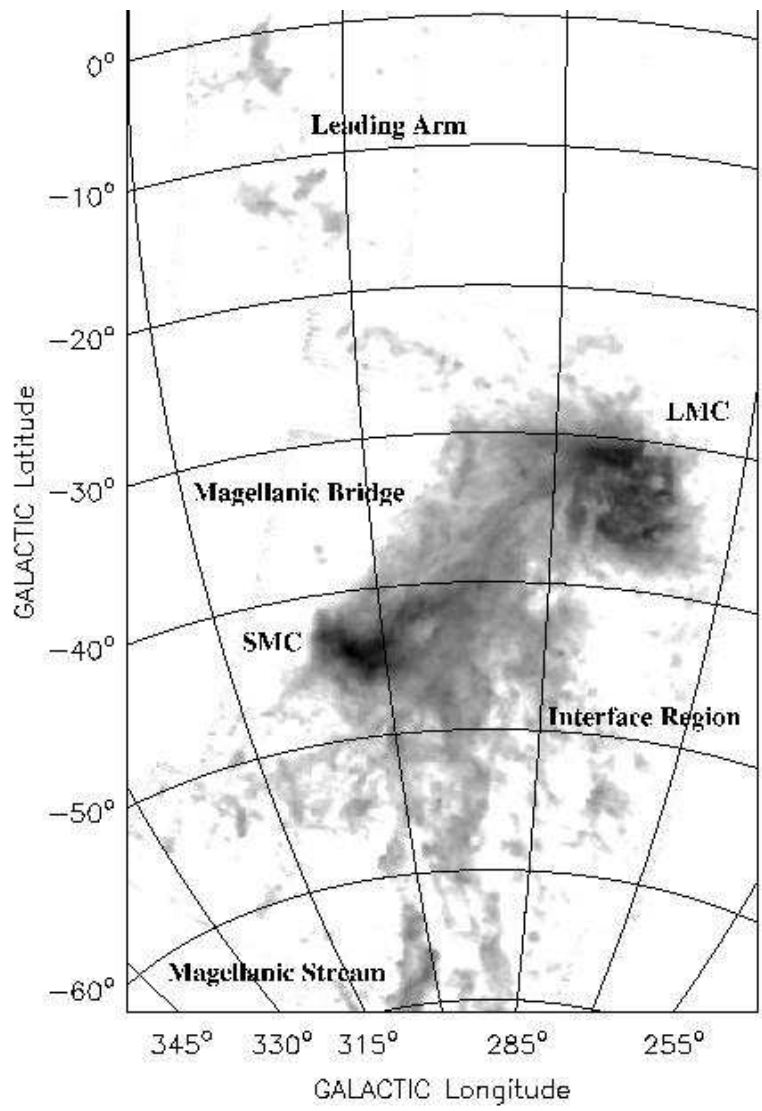}
 \vskip0.5cm
\caption{LMC \& SMC: an interacting system of dwarf galaxies (Br\" uns et al. 2004).}
\label{fig1}
\end{figure}

The high resolution HI surveys of the LMC (Kim et al. 1998, 1999), of
the SMC (Staveley-Smith et al. 1997, Stanimirovich et al. 1999), of the
Magellanic bridge region (Muller et al. 2003) and of all the system
(B\" uns et al. 2004) have been preformed with the Australia Telescope
Compact Array and with the Parkes radiotelescope.    
The following large-scale features are distinguished on the (l, b)
integrated HI intensity maps: LMC, SMC, Magellanic Bridge  joining the
two clouds, Magellanic Stream  starting at the SMC and following
an almost polar plane passing less than about 10$^\circ $ of the galactic
south pole and stretching more than 100$^\circ $, the Leading Arms  -
the HI gas preceding the motion of the clouds, and the Interface
Region  - the HI between Magellanic Bridge and Magellanic Stream 
(Fig. \ref{fig1}). All
these features are also distinguished on the average radial velocity
maps. The radial velocity changes smoothly from
$RV_{LSR} = - 400\  km\  s^{-1}$ at the end of the Magellanic Bridge
to $+ 100\  km\  s^{-1}$ at the SMC,
to $+ 240\  km\  s^{-1}$ at the LMC, while the Leading Arm does not show 
a clear
gradient. Deprojection correcting for the solar motion reduces the
velocity difference between the LMC and SMC from $\Delta
RV_{LSR} = 123\  km\  s^{-1}$ to $\Delta RV_{GSR} = 67\  km\  s^{-1}$. 
A further
correction taking into account motions of the LMC relative to the MW
reduces the difference to $\Delta RV_{LMCSR} = 10\  km\  s^{-1}$.
The HI average radial velocities show that the encounter between LMC and
SMC happens at a small velocity not much larger that $10\  km\  s^{-1}$,
which makes the interaction rather long, giving the time to the gas to
flow away from its parent cloud and form  the observed features.   

The LMC HI disk seems to be compressed at the side opposite to the
SMC, and the LMC shows rotation almost perpendicular to SMC - LMC
direction. The LMC disk  has a diameter of about 7.3 kpc with a
rotation curve rising  rapidly to $55\ km\ s^{-1}$ in the inner 1.5
kpc, more smoothly to a $63\ km\ s^{-1}$ peak at 2.4 kpc and declining
thereafter (Kim et al., 1998). The SMC also shows rotation: SMC disk
includes the bar-like
feature about 4 kpc in extent, with a velocity gradient of about $100\
km\ s^{-1}$ (Stanimirovi\' c  et al. 2004). 
The rotation curve rises to about $60\ km\ s^{-1}$ up to the turnover
radius of 3 kpc. The velocity dispersion is high along the high column
density axis of the Magellanic Bridge. LMC, Magellanic Bridge and SMC
show similar velocity fields,
which can result from the rotation of all the three partners with the
same orientation along the axis LMC - SMC. The velocity field is
rather broken at the southern part (in galactic coordinates) of the
SMC and of the Magellanic Bridge. There, the Magellanic Stream and
Interface Region start. At the other end, in
front of the LMC, the Leading Arm can be split into three parts. Individual
features in the Leading Arm show head-tail structure with the orientation along
the direction of possible space motion of the LMC as it is given by
Kroupa \& Bastian (1997) and van der Marel (2001).   

The HI observation should be complemented with the studies of 
distribution of planetary nebulae (Dopita et al. 1985), of carbon stars
(Kunkel et al. 2000), and of Cepheids (Groenewegen 2000), and with the
near-infrared star counts from the Two Micron All Sky Survey (2MASS) and
the Deep Near-Infrared Southern Sky Survey (DENIS). van der
Marel (2001) shows that LMC disk is not circular at larger radii, it is
elongated in the direction of the Galactic centre, suggesting the
influence of the tidal forces of the Milky Way. The data should be
compared to models of the interaction in attempts not to be in contradiction.

The N-body simulations modeling the gravitational interaction of the
SMC with the LMC and Milky Way (Gardiner \& Noguchi, 1996) show that the last
two close encounters between the interaction partners, 1.5 and 0.2 Gyr
ago, are able to explain many of the observed structures. Magellanic
Stream and Leading Arm have been created as a consequence of the former
close encounter, 1.5 Gyr ago. The Magellanic Bridge and the Interface
Region    
\noindent
have been formed later during the last close
encounter between LMC and SMC 0.2 Gyr ago. The discovery of the Leading
Arm stressed the importance of gravitational tides and questioned the role
of ram pressure stripping (Putman et al. 1998).  
However, the separation between gaseous and stellar features,
absence of stars in tidal Magellanic Bridge and Magellanic Stream, 
filamentary structures in the SMC and head-tail structures in the
Leading Arm and in the Magellanic Stream show that also hydrodynamical
forces like ram pressure
gas stripping have to act there. Most of the gas in Magellanic Bridge
and Leading Arm is coming from the
SMC. The preencouter SMC, more that 1.5 Gyr ago, must have a gas disk of
about 10 kpc in diameter (Stanimirovi\' c 2004), which shrunk forming
the Magellanic Stream and Leading Arm 1.5 Gyr ago and Magellanic
Bridge and Interface Region 0.2 Gyr ago. However, still a
substantial amount of angular momentum of the original disk is left
corresponding to other
simutions of dwarf galaxies (Mayer et al. 2001) demonstrating that the
tidal stripping removes the angular momentum rather slowly, at the
timescale of 10 Gyr. Consequently, the recent two encounters have not
been able to remove a substantial part of the original angular
momentum making possible to see the rotation of the SMC at present
times. The present rotation curve of the SMC shows that the dynamical
mass within 4 kpc is about $2.4\ 10^9 M_{\odot }$ three-quarters of
which is stellar. The dark matter is not needed for an explanation of
the rotation speeds in the SMC. 

Careful analysis of the HI distribution in the Magellanic Stream (Putman
et al. 2003) suggests that the Magellanic Bridge is older that assumed
above and the LMC and the SMC are bound together for at least two
orbits. The dual filaments emanating from the SMC and from the
Magellanic Bridge are of tidal origin and shaped by a small amount of
ram pressure. 

Star formation and energy feedback from young stars are included in
the so far most sophisticated N-body model of the LMC - SMC -
Milky Way encounter by Yoshizawa \& Noguchi (2003).  
This model agrees well with several observed features including the
Magellanic Stream, which apparently has tidal origin, it reproduces 
the presence of
young stars in the south-east wing of the SMC and it also reproduces
the acceleration in the star formation activity, which is due
to recent close encounters between the clouds. Some open and unsolved 
questions remain:
bimodal or many peak distribution in the main gaseous body of the SMC
and Magellanic Bridge remain to be interpreted, it probably originates in the
numerous expanding shells. The shells and supershells, if they trigger
star formation, also remain an open question.

The influence of other MW satellites is unclear. The tidal stream of
HI clouds connected to disruption of the Sagittarius dwarf galaxy on
the polar orbit aloud the Milky Way provides a possible explanation  
for the anomalous velocity distribution of HI clouds near the south galactic pole
(Putman et al. 2003). The possibility
of formation of the local group dwarf members including LMC and SMC
out of the Milky Way encounter with the M31 galaxy is discussed 
(Sawa \& Fujimoto, 2004). Another picture describes the interaction of the
Fornax-LeoI-LeoII-Sculptor-Sextans stream with the Magellanic Stream causing the gas
stripping from the Fornax (Dinescu et al. 2004). Kroupa et al. (2004)
propose the origin of the
whole Local Group with the local dwarf galaxies in a common great
circle.

\section{Star formation, energy and mass feedback}

Star formation is a complex process of the gravitational collapse and  
fragmentation, where the thermal and magnetic support competes  with 
supersonic shock waves and energy dissipation. The density increases by 20 
orders of magnitude from that of a molecular clouds core. The interplay 
between gravity, magnetic forces, hydrodynamical processes, radiative 
transfer and chemistry happens in a turbulent interstellar medium.     
Supersonic flows form sheets and filaments involving mass concentrations, 
which 
in some cases are bound by self-gravity. Some of them collapse 
forming single or binary stars, others disperse. 

There are different sources powering the interstellar structures.
On a small scale the pre-main sequence stellar winds and stellar radiation, 
on somewhat larger scale the main sequence stellar winds, and on even larger 
scale the supernovae. Young stars pump energy back to the interstellar medium, 
which influences the conditions for further star formation. The energy 
released by young OB associations compresses the ambient medium into shells 
and super-shells, which may collapse and trigger new star formation. The energy
feedback triggering shell collapse and further star formation is a 
self-regulating mechanism of the galaxy evolution.

\subsection{The observation of shells}

\begin{figure}
 \centering
 \includegraphics[angle=0,width=7.5cm]{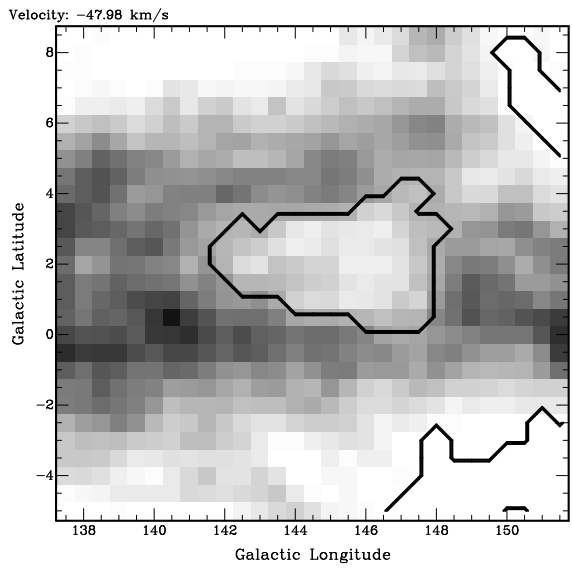}
 \includegraphics[angle=0,width=7.5cm]{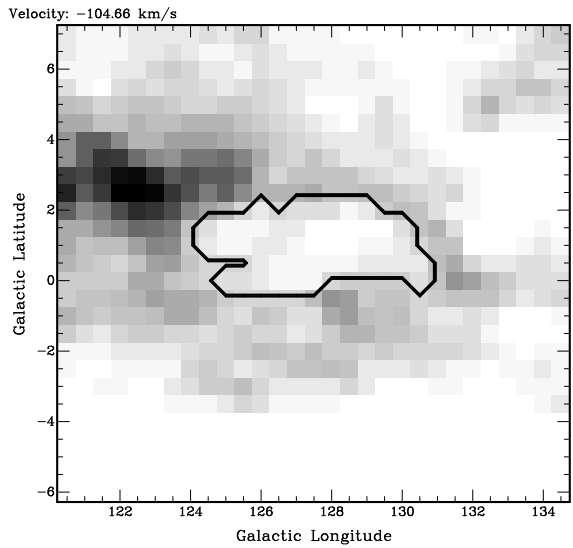}
 \caption{Re-identification of the shell GS 128+01-105 discovered by
          Heiles (1979) - upper frame, and a newly discovered shell 
           - lower frame.}
 \label{figyn}
\end{figure}

Shells and supershells and holes in the HI distribution have been discovered 
in the Milky Way by Heiles (1979, 1984), in M31 by Brinks and Bajaja (1986), 
in M33 by Deul and Hartog (1990), in LMC by Kim et al. (1999), in SMC by 
Stanimirovi\` c (1999), in HoII by Puche et at. (1992), 
in Ho I by Ott et al. (2001) 
and in IC 2574 by Walter and Brinks (1999).
Most probably they are created by an energy release from massive stars, 
however, an alternative explanations, infall of high velocity clouds 
(Tenorio-Tagle and Bodenheimer, 1988), or gamma ray bursts (Efremov et al., 
1998; Loeb and Perma, 1998) has been invoked in some cases. The majority 
of the 
observed shells is due to star formation (Ehlerov\' a and Palou\v s, 1996). 
In a new search by Ehlerov\' a and Palou\v s (2004), more than 600 shells 
have been identified in the Leiden-Dwingeloo HI survey 
of the Milky Way. 
In Fig. \ref{figyn} we show the re-identification of a shell
previously discovered by Heiles (1979) and a newly discovered shell.
The distribution of them in
the radial galactocentric direction and in the direction perpendicular to the 
galactic disk is similar to stellar distribution supporting the idea of 
a connection between massive stars and shells.     

\subsection{The collapse of shells}

\begin{figure}
\begin{center}
 \includegraphics[width=11cm,angle=0]{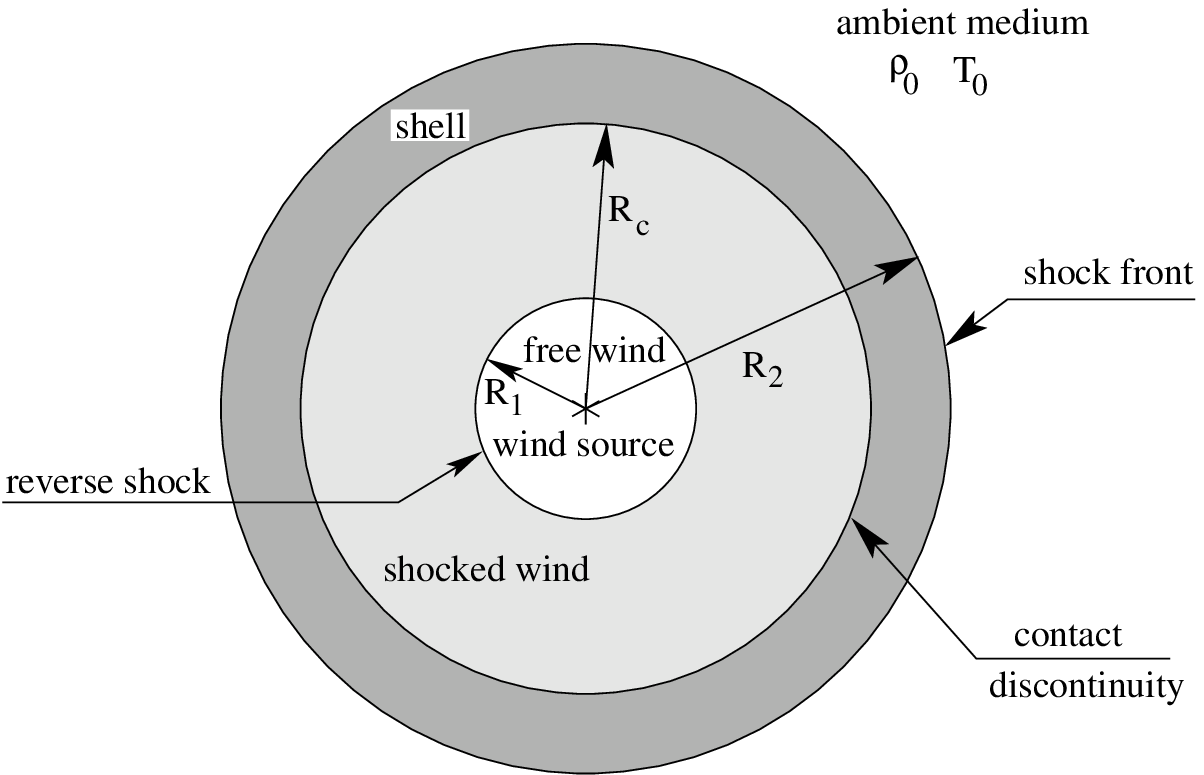}
\end{center}
\caption{Schematic representation of a supershell expanding around an OB
association.}
\label{shell}
\vspace*{0.5cm}
\end{figure}

We discuss the supersonic expansion of shells and sheets from regions
of localized deposition of energy and address the question if and when they
fragment and collapse due to gravitational instability. 
The energy input from an OB association creates a blast-wave
which propagates into the ambient medium (Ostriker \& McKee, 1988;
Bisnovatyi-Kogan \& Silich, 1995). The schematic representation of the
situation is shown in Fig. \ref{shell}. After the initial fast expansion the
mass accumulated in the shell cools and collapses to a thin
structure, which is approximated as infinitesimally thin layer
surrounding the hot medium inside.  Neglecting the external pressure
and assuming the constant energy input $L$, the self-similar solution
for radius $R$, expansion velocity $V$ and column density $\Sigma_{sh} $ is
(Castor et al. 1975; Ehlerov\' a \& Palou\v s, 2002):

\begin{eqnarray}
    R(t) & = 53.1 \times 
    \left ({L \over 10^{51}~\mathrm{erg~Myr}^{-1}}\right )^{1 \over 5}
    \times
    \left ({\mu \over 1.3}{n \over \mathrm{cm}^{-3}}\right )^{-{1 \over 5}}
    \times
    \left ({t\over \mathrm{Myr}}\right )^{3\over 5} \mathrm{pc} \\
   V(t) & = 31.2 \times 
    \left ({L \over 10^{51}~\mathrm{erg~Myr}^{-1}}\right )^{1 \over 5} \times
    \left ({\mu \over 1.3}{n \over \mathrm{cm}^{-3}}\right )^{-{1 \over 5}}
    \times
   \left ({t\over \mathrm{Myr}}\right )^{-{2\over 5}} \mathrm{km s}^{-1}\\
    \Sigma(t)_{\mathrm{sh}} & = 0.564 \times 
    \left ({L \over 10^{51}~\mathrm{erg~Myr}^{-1}}\right )^{1 \over 5} \times
    \left ({\mu \over 1.3}{n \over \mathrm{cm}^{-3}}\right )^{4 \over 5}
    \times
    \left ({t\over \mathrm{Myr}}\right )^{3\over 5}~\mathrm{M_{\odot}
    \mathrm{pc}^{-2}},
\end{eqnarray}
where $n$ is the density, $\mu $ is the mean atomic weight of
the ambient medium and $t$ is the time since the beginning of an expansion.

The linear analysis of hydrodynamical equations including  
perturbations on the surface of the shells
has been performed by Elmegreen (1994) and W\" unsch \& Palou\v s
(2001). The fastest growing mode is: 
\begin{figure}
\begin{center}
\includegraphics[width=9cm,angle=0]{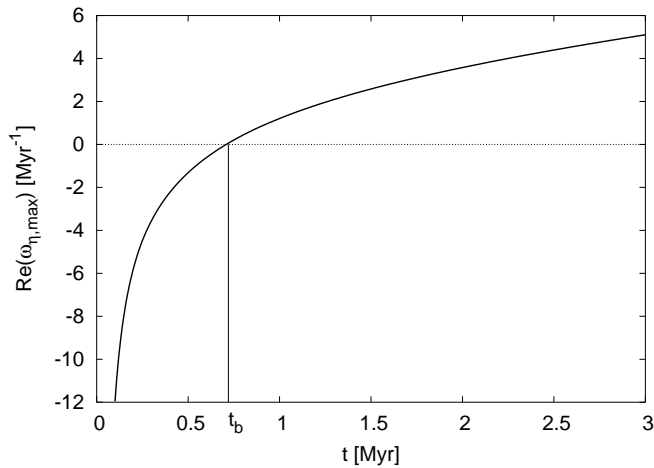}
\end{center}
\caption{The fastest growing mode.}
\label{omega}
\end{figure}

\begin{equation}
  \omega = -{3V \over R} + \sqrt{{V^2 \over R^2} +
  \left ({\pi G \Sigma_{sh}
  \over c_{sh}}\right)^2},
\end{equation}
where $c_{sh}$ is the sound speed inside of the expanding shell.

In Fig. \ref{omega}. we give the time evolution of the fastest mode. 
At early times, for 
$t < t_b$, the shell is stable. $t_b$ is the time, 
when the fastest mode starts to be unstable:
\begin{equation}
   t_{\mathrm{b}} = 28.8 \times 
    \left ({c_{\mathrm{sh}} \over \mathrm{km~s}^{-1}}\right )^{{5 \over 8}}
    \times
    \left ({L \over 10^{51}~\mathrm{erg~Myr}^{-1}}\right )^{-{1 \over 8}}
    \times
    \left ({\mu \over 1.3}{n \over \mathrm{cm}^{-3}}\right )^{-{1\over 2}}
    \mathrm{Myr}.
\end{equation}
Later, for $t > t_b$, when the expansion slows
down and reduces the stretching, which acts against gravity, and when the
shell column density increases, the shell starts to be gravitationally
unstable. For ambient densities similar to values in the 
solar vicinity, $n \sim 10^{-1} - 10^2$, $t_b$ is a few $10^7$ yr, 
which means that the gravitational instability is rather slow compared to 
turbulent collision times and galactic differential rotation. 
$t_b$ is much smaller in high density medium of GMC and dense cores, 
where it is $\sim 10^4$ yr only.
Thus the shell gravitational instability is particularly important inside 
the GMCs.      

\begin{figure}[h]
\begin{center}
 \includegraphics[angle=0,width=9cm]{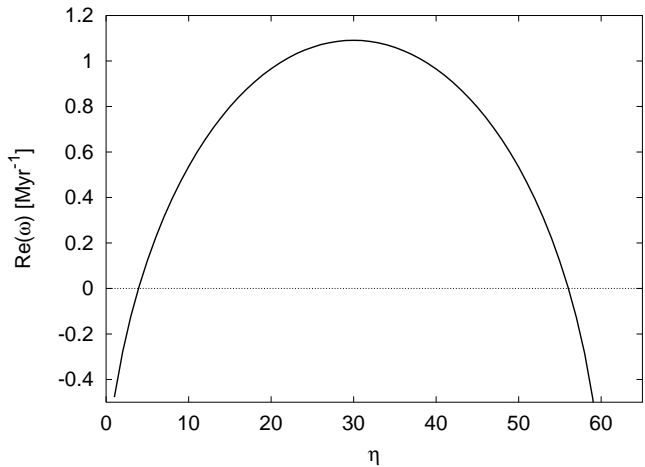}
\end{center}
\caption{The dispersion relation}
\label{disper}
\end{figure}

The dispersion relation of the shell gravitational instability is: 
\begin{equation}
\omega (\eta , t) = -{3 V \over R} + \sqrt {{V^2 \over R^2} - {c_{sh}^2
\eta ^2 \over R^2} + {2 \pi G \Sigma_{sh} \eta \over  R}},
\end{equation}
where $\eta $ is the dimensionless wavenumber and $\lambda $ is the 
wavelength of the perturbation: $\eta = 2 \pi R / \lambda $.
It is shown in Fig. \ref{disper}: it gives the wavelength interval of unstable 
perturbations.

The resulting number of fragments is inversely proportional to the
fragment growth time $t_{growth} = {2 \pi \over \omega (\eta, t) }$. 
Rapidly growing fragments are more
frequent in the final mass spectrum than the slowly growing fragments.
\vskip 0.5cm
\noindent
Thus the number of fragments in a given volume of radius $R$ 
is
\begin{equation}
N = \omega {R^3 \over (\lambda /4)^3}. 
\end{equation}
\begin{figure}[h]
\begin{center}
 \includegraphics[angle=270,width=10cm]{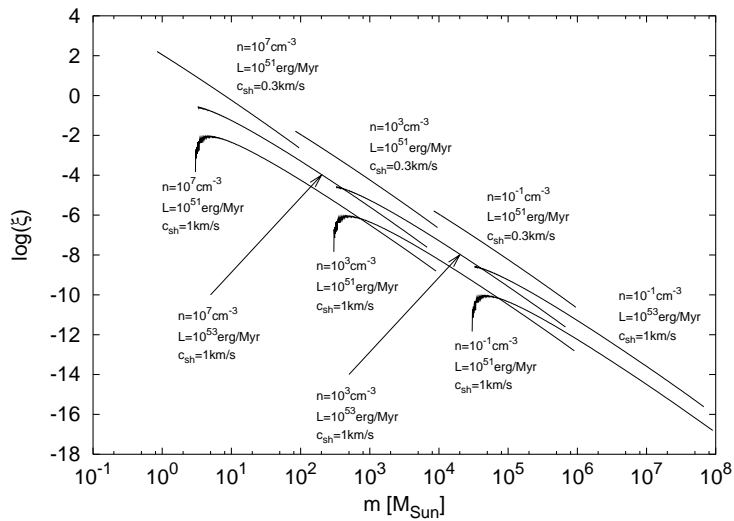}
\end{center}
\vskip0.5cm
\caption{The mass spectrum of fragments of an expanding shell at time
 $5 t_{\mathrm{b}}$.}
\label{spectrum}
\end{figure}
A fragment with the wavelength $\lambda $ has the mass
\begin{equation}
m = {4 \over 3} \pi (\lambda /4)^3 \rho.
\end{equation}
We  derive the mass spectrum $\xi (m) = {dN \over
dm}$:
\begin{equation}
\xi (m) = - {4 \over 3}\  \pi\  R^3\ \rho \ \omega \ m^{-2}.
\end{equation}

If the dispersion relation $\omega(\eta )$ were a constant, the
slope of the mass spectrum  $\xi (m) \propto m^{-\alpha }$
would be exactly approximated with a power
law slope  $\alpha $= 2. But in our case $\omega $ is not only the function of
$\eta $ but also of the time $t$. 

We assume that $t_{growth}$ for a given $\eta $ is inversely proportional 
to the
average value  of $\omega $ for this $\eta $ since the time $t_b (\eta )$ 
when a given mode starts to be unstable. 
The time average $\bar\omega $ for a given $\eta $ is calculated using the equation:
\begin{equation}
\bar\omega (\eta ) = {\int_{t_{\mathrm{b}}(\eta )}^t \omega(\eta ,t') dt' 
\over t - t_b (\eta )}.
\end{equation}
The resulting mass spectra for different values of $n$, $c_{sh}$ and
$L$, as they have been derived using the thin-shell approximation (4) -
(6), are shown in Fig. \ref{spectrum}. The high mass parts are well approximated by
the power law with a slope $\alpha = 2.2 - 2.4$.

In low density medium, the collapse of expanding shells forms fragments
with masses comparable to GMC, however, the collapse time is rather
long, a few $10^7$ yr. In high density medium of GMC cores, the
collapse time is rather short, $\sim 10^4$ yr, the masses
of individual fragments are close to stellar and the initial mass
function of fragments has a power law slope close to the Salpeter (1955) value
-2.35. We conclude that the fragmentation of expanding
shells qualifies as a possible process triggering the star formation
in environments, where the density is high enough, or where it has been
increased due to mass accumulation.

\section{Formation of super-star clusters}

\begin{figure}
\includegraphics[width=11cm,angle=0]{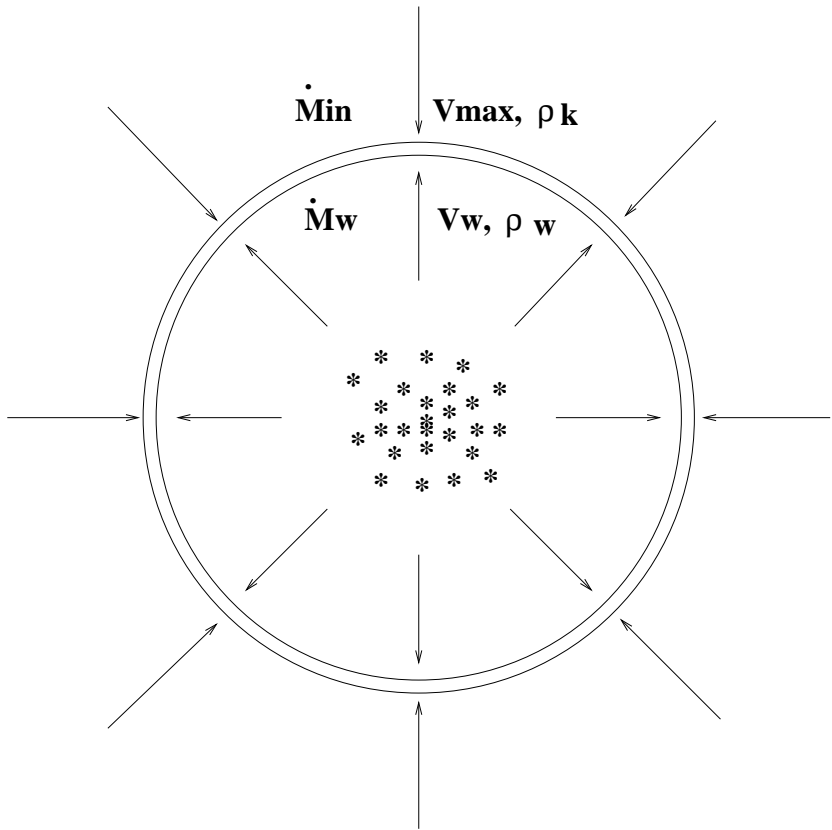}
\caption{The shell inside of a collapsing cloud}
\label{factory}
\end{figure}

Young and massive ($10^5 - 10^7 M_{\odot }$) super-star clusters are
observed along tidal arms and bridges of the colliding galaxies. A
model of a star forming factory is proposed by Tenorio-Tagle et
al. (2003).
This model invokes pressure-bounded, self-gravitating, isothermal
cloud (Ebert, 1955; Bonner, 1956), which becomes gravitationally
unstable when sufficiently compressed. The gravitational instability
allows the cloud to enter isothermal collapse. As the collapse
proceeds a first generation of stars is formed in the  center of the
cloud.  The mass
and energy feedback of the first generation of stars has an important
impact on the collapsing cloud. Stellar winds and supernovae compress
the infalling material forming a dense shell (see
Fig. \ref{factory}). The shell is able
to trap the ionizing radiation and winds of the first generation
of young stars. At the same time the shell fragments forming stars with a high
efficiency. Thus the cloud, which is compressed in the case of the
galaxy encounter, forms a new super-star cluster.

\section{Evolution of the star formation rate}

The investigation of
colors in Hubble deep fields (Madau et al., 1996; 1998;
Rowan-Robinson, 2003) provides extinction as a function of $z$: it has
been higher at $z = 0.5 - 1.5$ than locally, and lower at $z >
2$. Related models of the star formation history show the steep
decline of the star formation rate since $z = 1$, the present star
formation has in average  about an order of magnitude smaller rate
compared to the level at $z = 1 - 1.5$. Even deeper in the past the
correction for the dust extinction remains highly uncertain and a
conclusion on the evolution of the star formation rate for $z = 2 - 6$
is difficult.

This opens the question what drives the star formation. Bars in
the central parts of galaxies does not seem to change substantially
between $z = 1$ and present. They do trigger the star formation,
however, there has to be more partners in the game. The fuel - the gas
- is depleted in galaxies not only due to star formation, but
also since the ISM is removed from the galaxy disks with tides, with
the ICM
ram pressure stripping and with the star formation feedback. 
Number of galaxy interactions also decreases.
We conclude that the decline of the star formation is due to lower rate
of triggering from galaxy interactions in a combination with the
starvation, since the amount of gas in the star forming disks is
reduced by environmental effects and by the star formation itself.      

\subsection*{Acknowledgements}

I would like to express my thanks to Guillermo Tenorio-Tagle for comments to 
the early version of the text.  

\subsection*{References}

{\small

\bref
Abadi, M. G., Moore, B., \& Bower, R. G. 1999, MNRAS, 308, 947

\bref
Amram, P., Mercelin, M., Boulsteix, J., \& Le Coarer, E. 1992, A\&A, 266, 106

\bref
Amram, P., Mendes de Oliveira, C., Plana, H., Balkowski, C., Hernandez, C., 
Carignan, C., Cypriano, E. S., Sodr\' e, L. Jr., Gach, J. L., \& Boulsteix, J.
2004, ApJ, 612, L5

\bref
Bekki, K., \& Chiba, M. 2002, 566, 245

\bref
Bertschik, M., \& Burkert, A. 2003, RevMexAA, 17, 144

\bref
Bisnovatyi-Kogan, G. S. \& Silich, S. A. 1995, RvMP 67, 661

\bref 
Bonner, W. B. 1956, MNRAS, 116, 351

\bref
Bournaud, F., \& Combes, F. 2004, A\&A, submitted

\bref
Bournaud, F., Combes, F., \& Jog, J. 2004, A\&A, 418, L27

\bref
Bournaud, F., Duc, P.-A., Amram, P., Combes, F., \& Gach, J.-L. 2004, A\&A, 
425, 813

\bref
Brinks, E., \& Bajaja, E. 1986, A\&A, 169, 14

\bref
Br\" uns, C., Kerp, J., Staveley-Smith, L., Mebold, U., Haynes, R. F.,
Kalberla, P. M. W., Muller, E., \& Filipovic 2004, A\&A, submitted

\bref
Bureau, M., \& Carignan, C. 2002, AJ, 123, 1316

\bref
Burkert, A. \& Naab, T. 2003a, in Coevolution of Black Holes and Galaxies,
ed. L. C. Ho, Carnegie Obs.

\bref
Burkert, A. \& Naab, T. 2003b, in Galaxies and Chaos, ed. G. Contopoulos \&
N. Voglis, Springer

\bref
Buson, L. M., Bertola, F., Bressan, A., Burstein, D., \& Cappellari,
M. 2004, A\&A, 423, 965

\bref
Butcher, H., \& Oemler, A. 1978, ApJ, 219, 18

\bref
Butcher, H., \& Oemler, A. 1984, ApJ, 285, 426

\bref
Castor, J., McCray, R., \& Weaver, R. 1975, ApJ, 200, L107

\bref
Chitre, A., \& Jog, C. J. 2002, A\&A, 388, 407

\bref
Ciotti, L., Pellegrini, S., Renzini, A., \& D'Ercole, A. 1991, ApJ,
376, 380

\bref
Combes, F. 2004, IAU Symp. 217, 440

\bref
Couch, W. J., Barger, A. J., Smail, I., Ellis, R. S., \& Sharples,
R. M. 1998, ApJ, 497, 188

\bref
Deul, E. R., \& den Hartog, R. H. 1990, A\&A, 229, 362

\bref
Dinescu, D. I., Keeney, B. A., Majewski, S. R, \& Terrence,
G. M. 2004, AJ, 128, 687

\bref
Dopita, M. A., Ford, H. C., Lawrence, C. J., \& Webster, B. L. 1985,
ApJ, 296, 390

\bref
Dressler, A., Oemler, A., Couch, W. J., et al. 1997, ApJ, 490, 577

\bref
Duc, P.-A., Bournaud, F.,\& Masset, F. 2004, A\&A, 427, 803

\bref
Duc, P.-A., Brinks, E., Springel, V., et al. 2000, AJ, 10, 1238

\bref
Ebert, R. 1955, Z. Astrophys., 36, 222

\bref
Efremov, Yu. N., Elmegreen, B. G., \& Hodge, P. W. 1998, ApJ, 501, L163

\bref
Ehlerov\' a, S., \& Palou\v s, J. 1996, A\&A, 313, 478

\bref
Ehlerov\' a, S., \& Palou\v s, J. 2002, MNRAS, 330, 1022

\bref
Ehlerov\' a, S., \& Palou\v s, J. 2004, A\&A, submitted

\bref
Elmegreen, B. G. 1994, ApJ, 427, 384

\bref
Gao, L., Loeb, A., Peebles, P. J. E., White, S. D. M., \& Jenkins,
A. 2004, ApJ, 614, 17

\bref
Gardiner, L. T., \& Noguchi, M. 1996, MNRAS, 278, 191

\bref
Gnedin, O. Y. 2003, ApJ, 582, 141

\bref
Groenewegen, M. A. T. 2000, A\&A, 363, 901

\bref
Gunn, J. E., \& Gott III, J. R. 1972, ApJ, 176, 1

\bref
Heckman, T. M. 2003, RMxAC, 17, 47

\bref
Heiles, C. 1979, ApJ, 229, 533

\bref
Heiles, C. 1983, ApJS, 55, 585

\bref
Hibbard, J. E., van der Hulst, J. M., Barnes, J. M., \& Rich, R. M. 2001,
AJ, 122, 2969

\bref
J\' achym, P. 2004, in preparation

\bref
Jungwiert, B., Combes, F., \& Palou\v s, J. 2001, A\&A, 376, 85

\bref
Jungwiert, B., Combes, F., \& Palou\v s, J. 2004, in preparation

\bref
Keel, W., Owen, F., Ledlow, M., \& Wang, D. 2004, AAS, 203, 4705

\bref
Kenney, J. D. P., \& Koopmann, R. A. 1999, AJ, 117, 181

\bref 
Kim, S., Staveley-Smith, L., Dopita, M. A., Freeman, K. C., Sault,
R. J., Kesteven, M. J., \& McConnell, D. 1998, ApJ, 503, 674

\bref
Kim, S., Dopita, M. A., Staveley-Smith, L., \& Bessell, M. S. 1999,
AJ, 118, 2797

\bref
Koopmann, R. A., \& Kenney, J. D. P. 2004, ApJ, 613, 851

\bref
Koopmann, R. A., \& Kenney, J. D. P. 2004, ApJ, 613, 866

\bref
Kroupa, P., \& Bastian, U. 1997, New A, 2, 77

\bref
Kroupa, P., Theis, Ch., \& Boily, C. M. 2004, A\&A, submitted

\bref
Kunkel, W. E., Demers, S., \& Irwin, M. J. 2000, AJ, 119, 2789

\bref
Li, Y., Mac Low, M.-M., \& Klessen, R. S. 2004, ApJ, L29

\bref
Loeb, A., \& Perma, R. 1998, ApJ, 503, L35

\bref
Madau, P., Ferguson, H. C., Dickinson, M. E., Giavalisco, M., Steidel,
Ch. C., \& Fruchter, A. 1996, MNRAS, 283, 1388

\bref
Madau, P., Pozzetti, L., \& Dickinson, M. 1998, ApJ, 498, 106

\bref
Mayer, L., Governato, F., Colpi, M., Moore, B., Quinn, T., Wadsley,
J., Stadel, J., \& Lake, G. 2001, ApJ, 559, 754

\bref
Mihos, J.Ch. 2004, IAU Symp. 217, eds P.-A. Duc, J. Braine and E. Brinks,
p. 390

\bref 
Mirabel, I. F., Dottori, H., \& Lutz, D. 1992, A\&A, 256, L19

\bref
Moore, B. 2003, in Clusters of Galaxies: Probes of Cosmological
Structure and Galaxy Evolution, ed. J. S. Mulchaey, A. Dressler, and
A. Oemler, CUP

\bref
Moore, B., Katz, N., \& Lake, G. 1996, ApJ, 457, 455

\bref
Moore, B., Katz, N., Lake, G., Dressler, A., \& Oemler, A. Jr. 1995,
Nature, 379, 613

\bref
Muller, E. Staveley-Smith, L., Zealey, W., \& Stanimirovi\' c,
S. 2003, MNRAS, 339, 105

\bref
Naab, T., \& Burkert, A. 2003, ApJ, 597, 893

\bref
Nulsen, P. 1982, MNRAS, 198, 1007

\bref
Ostriker, J. P., \& McKee, C. F. 1988, RvMP 60, 1

\bref
Ostriker, J., \& Peebles, J. A. 1973, ApJ, 186, 467   

\bref
Ott, J., Walter, F., Brinks, E., van Dyk, S. D., Klein, U. 2001, AJ, 122, 3070

\bref
Pettini, M., Rix, S. A., Steidel, Ch. C., Shapley, A. E., \&
Adelberger, K. L. 2003, IAU Symp. 212, 671

\bref
Puche, D., Westpfhal, D., Brinks, E., \& Roy, J. 1992, AJ, 103, 1841

\bref
Putman, M. E., Staveley-Smith, L., Freeman, K. C., Gibson, B. K., \&
Barnes, D. G. 2003, ApJ, 586, 170

\bref
Putman, M. E., Gibson, B. K., Staveley-Smith, L. + 23 co-authors 1998, 
Nature, 394, 752

\bref
Renzini, A. 1997, ApJ, 488, 35

\bref
Rezini, A. 2004, in Clusters of Galaxies: Probes of Cosmological Structure and
Galaxy Evolution, Cambridge Univ. Press, eds J. S. Mulchaey, A. Dressler, \&
A. Oemler, p. 261

\bref
Rowan-Robinson, M. 2003, MNRAS, 345, 819

\bref
Salpeter, E. E., 1955, ApJ, 121, 161   

\bref
Sawa, T., \& Fujimoto, M. 2004, astro-ph/0404547

\bref
Shen, J., \& Selwood,J. 2004, ApJ, 604, 614

\bref
Stanimirovi\' c, S., Staveley-Smith, L., Dickey, J. M., Sault, R. J.,
\& Snowden, S. L. 1999, MNRAS 302, 417

\bref
Stanimirovi\' c, S., Staveley-Smith, L., \& Jones, P. A. 2004, ApJ,
604, 176

\bref
Staveley-Smith, L., Sault, R. J., Hatzidimitriou, D., Kesteven, M. J.,
\& McConnell, D. 1997, MNRAS, 289, 225

\bref
Tenorio-Tagle, G., \& Bodenheimer, P. 1988, ARA\&A, 26, 145

\bref
Tenorio-Tagle, G., Palou\v s, J., Silich, S., Medina-Tanco, G. A., \&
Mu\~ noz-Tu\~ non 2003, A\&A, 411, 397

\bref
Toomre, A. 1964, ApJ, 139, 1217

\bref
Toomre, A. 1977, ARA\&A, 15, 437 

\bref
Toomre, A. \& Toomre, J. 1972, ApJ, 178, 623

\bref
van der Marel, R. P. 2001, AJ, 122, 1827

\bref
van Zee, L., Skillman, E. D., \& Haynes, M. P. 2004, AJ 128, 121

\bref
van Gorkom, J. H. 2004, in Clusters of Galaxies: Probes of
Cosmological Structure and Galaxy Evolution, Cambridge University Press,
Carnegie Observatories Astrophysics Series, eds. J. S. Mulchaey,
A. Dressler, and A. Oemler, p. 306
 
\bref
Vollmer, B., Cayatte, V., Balkowski, C., \& Duschl, W. 2001, ApJ, 561, 708

\bref
Walter, F., \& Brinks, E. 1999, AJ, 118, 273

\bref
W\" unsch, R., \& Palou\v s, J. 2001, A\&A, 374, 746

\bref
Yoshizawa, A., \& Noguchi, M. 2003, MNRAS 339, 1135
}

\vfill

\end{document}